\documentstyle[epsf]{mn}
{}
\begin{document}
\title[A new magnetic white dwarf : PG2329+267]
    {A new magnetic white dwarf : PG2329+267}
 
\author[C.Moran, T.R.Marsh and V.S.Dhillon]
{C. Moran$^{1}$, T. R. Marsh$^{1}$, and V. S. Dhillon$^{2}$\\
$^{1}$University of Southampton, Department of Physics, Highfield, Southampton SO17 1BJ. \\
$^{2}$Royal Greenwich Observatory, Madingley Road, Cambridge, CB3 0EZ.\\
}
 
\date{Accepted 28 April 1998\\
Received : 15 December 1997\\
In original form 29 October 1997}
 
\maketitle
 
\begin{abstract} 

We have discovered that the white dwarf PG 2329+267 is magnetic, and
assuming a centered dipole structure, has a dipole magnetic field
strength of approximately 2.3MG. This makes it one of only
approximately 4\% of isolated white dwarfs with a detectable magnetic
field. Linear Zeeman splitting as well as quadratic Zeeman shifts are
evident in the hydrogen Balmer sequence and circular
spectropolarimetry reveals $\sim$10\% circular polarisation in the two
displaced $\sigma$ components of H$_{\alpha}$. We suggest from
comparison with spectra of white dwarfs of known mass that PG 2329+267
is more massive than typical isolated white dwarfs, in agreement with
the hypothesis that magnetic white dwarfs evolve from magnetic
chemically peculiar Ap and Bp type main sequence stars.

\end{abstract}


\section{Introduction}

The possibility that white dwarfs may possess large magnetic fields was
first suggested in 1947 (Blackett 1947), however it was not until 1970
that the first detection was made (Kemp et al 1970). Since then 43
magnetic white dwarfs have been found, with field strengths ranging
from $\sim$0.1 to $\sim$ 1000 MG. The vast majority (96\%) of white
dwarfs have as yet shown no sign of magnetic fields (Schmidt \& Smith
1995) though this percentage may drop as surveys for magnetic fields
are extended to lower field strengths. Serendipitous discoveries made
during spectroscopic studies may also add to the number of known
magnetic systems. What the actual percentage of magnetic white dwarfs
is amongst the complete population, and what the distribution of field
strengths amongst this set is, remains unclear.

A more complete knowledge of the magnetic properties of these stellar
remnants, particularly at low (sub MG) magnetic field strengths, may
allow us to deduce the role played by magnetic fields throughout the
life time of the progenitor stars, as the magnetic fields found in
white dwarfs are believed to be fossil fields preserved from earlier
stages of stellar evolution. There is no known process for the
generation of large scale magnetic fields during the degenerate phase
of stellar evolution and so they are likely to be amplified versions
of the fields which permeated their parent stars. White dwarfs with
magnetic field strengths $>$1MG may be explained by evolution from
chemically peculiar, magnetic Ap and Bp stars (Angel et al 1981),
which have detectable magnetic fields from 100 to 10,000 G. White
dwarfs with weaker magnetic fields would require their main sequence
progenitors to have fields of only a few Gauss, below the current
observational limits. This theory is supported by the similar space
density of magnetic degenerates and the expected distribution of the
remnants of magnetic main sequence stars (Sion et al 1988), as well as
the observed tendency for magnetic white dwarfs to be more massive
than non magnetic white dwarfs due to their proposed evolution from
more massive progenitors.

The presence of a magnetic field has several detectable effects upon
the spectrum of a white dwarf. For magnetic field strengths between 1
and 20 MG the linear Zeeman effect produces a distinctive triplet
pattern for each absorption feature. Both the upper and lower atomic
levels split into three energetically equidistant sub-levels. This
allows transitions between the upper and lower levels to occur at
three different energies. The wavelength of the central $\pi$
component is unaffected by the presence of the magnetic field, however
the two $\sigma$ components are shifted, one to a longer
($\sigma^{-}$) and one to a shorter wavelength ($\sigma^{+}$). The
degree of this separation ($\pi - \sigma$) is determined by the
strength of the magnetic field (Landstreet 1994) according to,

\[ \Delta\lambda_{L} \simeq 4.7 \times 10^{-7}\lambda^{2}B_{s} \hspace{1cm}(1) \]

\noindent where $\lambda$ is measured in Angstroms and the average
magnetic field strength over the visible hemisphere of the white
dwarf, B$_{s}$ is measured in MG.

Above about 20MG the quadratic Zeeman effect dominates over the linear
effect and the spectra become more and more complicated. Even at lower
magnetic field strengths the quadratic Zeeman effect is noticeable as a
blue shift in the wavelength of all the lines in the spectra. The size
of the wavelength shift $\Delta\lambda_{Q}$, given by equation 2, is
different for each line in the Balmer series, with the higher lines
being shifted far more than H$_{\alpha}$ (Preston 1970),

\[ \Delta\lambda_{Q} \simeq -5 \times 10^{-11}\lambda^{2}n^{4}B_{s}^{2} \hspace{1cm}(2) \]

\noindent where n is the principle quantum number of the upper level
of the transition, so for the Balmer series n = 3 for H$_{\alpha}$ and
n = 8 for H$_{\zeta}$. This simple expression is based on perturbation
theory and will break down for high n values, even at quite modest
field strengths (Surmelian \& O'Connell 1974).
  
The circular polarisation of the light can also be used to measure the
magnetic field strength of a white dwarf. Even for weak magnetic
fields ($<$1MG), where the Zeeman splitting is not obvious due to the
large Stark broadening of white dwarf spectral features, the line
profile is still a superposition of the unshifted $\pi$ component and
the two shifted $\sigma$ components. In a longitudinal magnetic field
the two $\sigma$ components have opposite circular polarisations and
hence even though the net circular polarisation of the line is zero,
the offset $\sigma$ components produce a distinctive S shaped feature
in the circular polarisation spectrum. The percentage of circularly
polarisation (V$_{\%}$) is proportional to the longitudinal magnetic
field strength B$_{e}$ and the normalised flux gradient of the zero
field $\pi$ line, as shown below, where $I_{\lambda}$ is the flux.

\[V_{\%}(\lambda) = 1.1 B_{e} \left(\frac{\lambda}{4861}\right)^{2}
\frac{1}{I_{\lambda}} \frac{dI_{\lambda}}{d\lambda} \hspace{1cm}(3) \]

Hence by measuring the degree of circular polarisation we can
calculate B$_{e}$, the mean longitudinal magnetic field strength over
the visible hemisphere of the white dwarf.

\section{Observations}

In November 1995 we observed the white dwarf PG 2329+267 as part of a
spectroscopic survey to determine the masses of DA (hydrogen
dominated) white dwarfs. It was immediately clear from the
characteristic Zeeman splitting of the Balmer lines that PG 2329+267
was magnetic. Follow up observations in January 1996, using circular
spectropolarimetry, confirmed the existence of a magnetic field.

The initial discovery of the magnetic nature of PG 2329+267 was made on
23 November 1995 using the IDS spectrograph and 235 mm camera on the
INT (2.5-m Isaac Newton Telescope), La Palma. The spectra cover 3682
to 5300 \AA \  at a FWHM resolution of 2.3 \AA.

We conducted follow up observations on 12 January 1996. We used the
spectropolarimeter with the blue arm of ISIS on the WHT (4.2-m William
Hershel Telescope), La Palma. The set up consisted of a quarter
waveplate to convert the circular polarisation into its two orthogonal
components, a dekker to separate star and sky spectra, the
spectrograph slit and then a calcite block to separate the two linear
polarisations. Each observation yielded two spectra of the object,
corresponding to the two rays split by the calcite block. In principle
the intensity difference of these two spectra could yield the
percentage of circular polarisation, however this would ignore any
differences in the response of the spectrograph and the detector
between the o and e rays. To account for any differences in
instrumental response, the quarter wave plate was rotated by 90
degrees and a second set of observations were made. The rotation of
the quarter wave plate resulted in the reversal of the paths of the
two rays. Hence by comparing the two exposures we could remove the
instrumental response. We observed H$_{\alpha}$, with the spectra
covering 6362 - 6769 \AA \ at a FWHM resolution of 0.7 \AA.

\begin{figure*} 
\begin{center}
\epsfxsize 0.99\hsize
\leavevmode
\epsffile{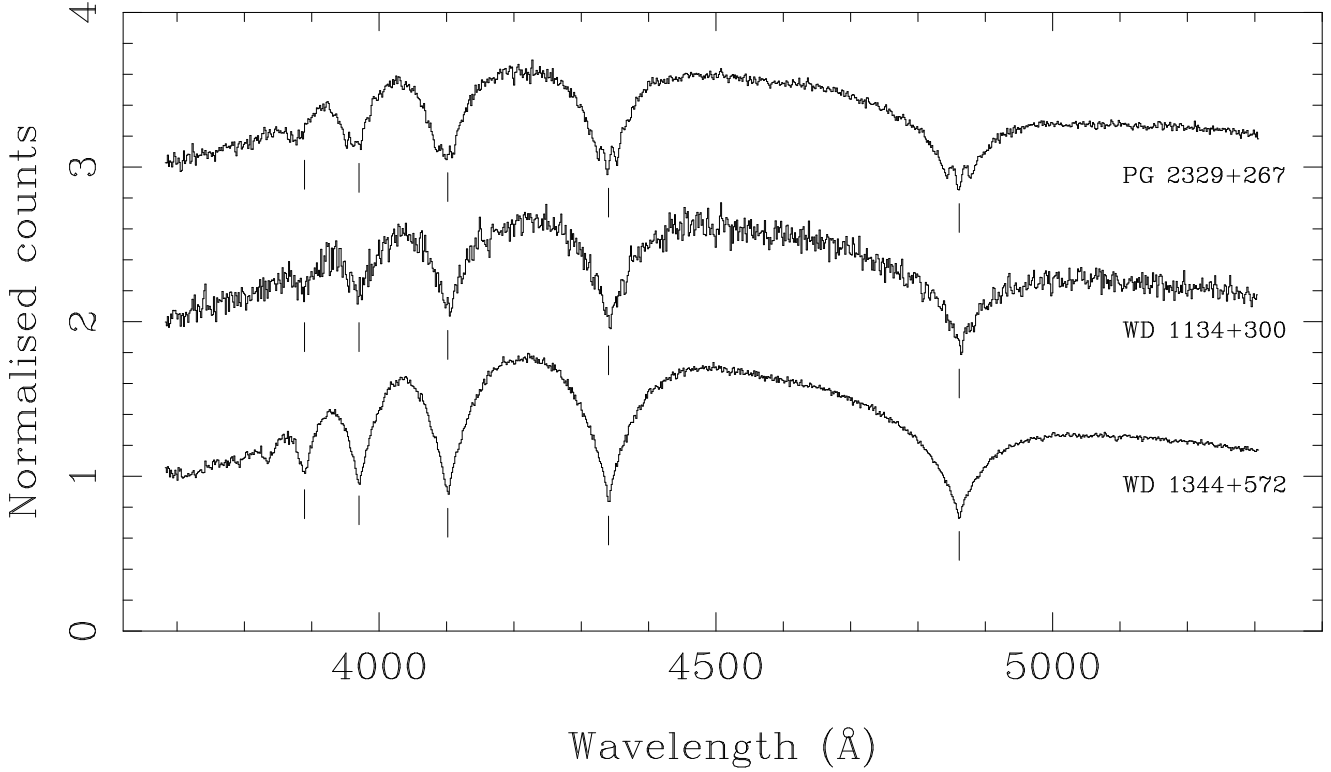}
\end{center}
\caption{The spectrum of PG 2329+267 (top) taken with the INT clearly shows Zeeman splitting of the hydrogen Balmer lines due to the presence of a magnetic field. The spectra of two non-magnetic white dwarfs are offset below for comparison. The vertical lines are placed at the rest wavelengths of the Balmer lines and aid detection of the quadratic zeeman shift in the spectrum of PG 2329+267.}\end{figure*}  

\begin{figure*} 
\begin{center}
\epsfxsize 0.99\hsize
\leavevmode
\epsffile{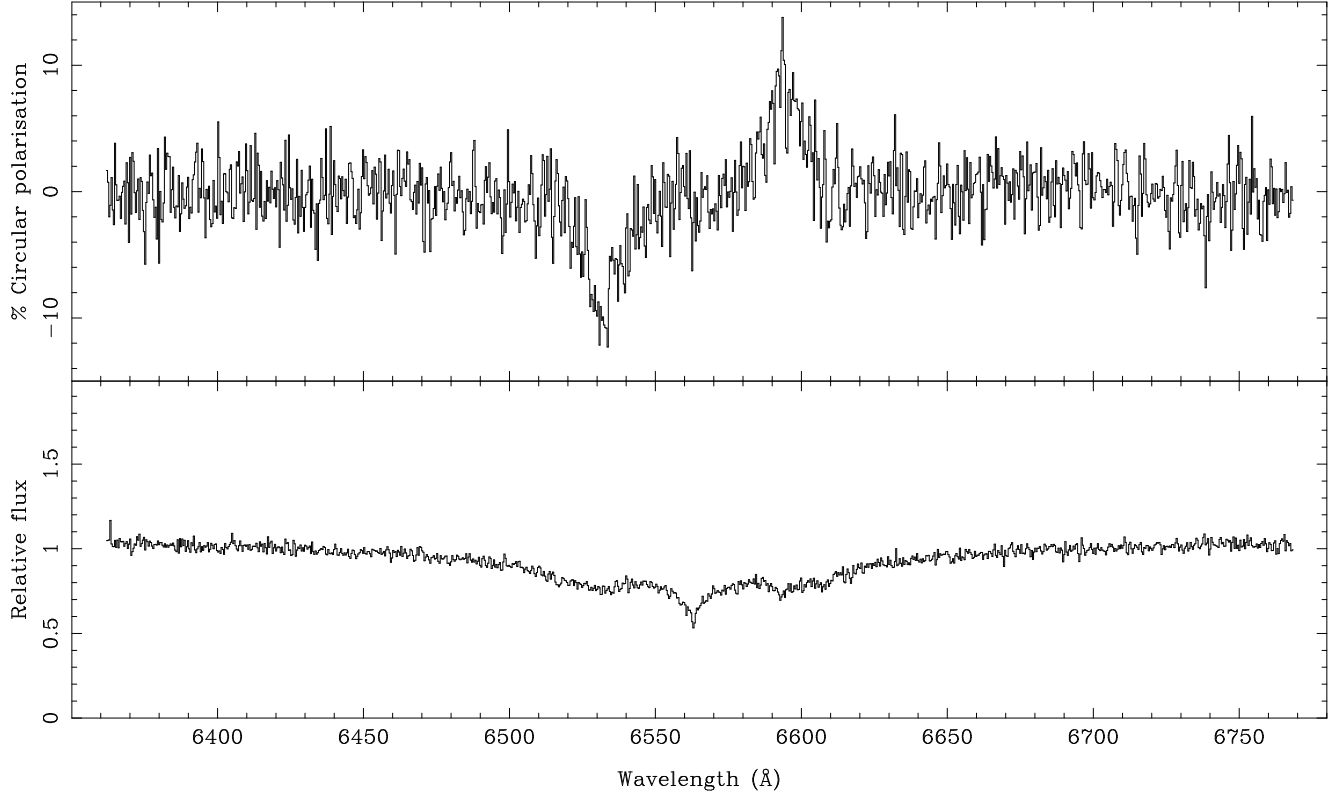}
\end{center}
\caption{The lower panel shows the normalised spectrum of H$_{\alpha}$
in PG2329+267. The top panel shows the percentage of circularly
polarised light present in the spectrum.}
\end{figure*} 

\begin{table*}
 \centering
 \caption{Linear and quadratic Zeeman features}
 \begin{tabular}{cccccc}
Line & B$_{s}$ MG (Linear) & $\lambda_{\pi}$ \AA & Quad shift \AA &
B$_{s}$ MG (Quad) & Corrected B$_{s}$ MG (Quad) \\
H$_{\alpha}$ & 1.47 $\pm$ 0.07 & 6563.04 $\pm$ 0.10 & 0.28 $\pm$ 0.10 & -- & -- \\
H$_{\beta}$ & 1.57 $\pm$ 0.15 & 4860.55 $\pm$ 0.40 & -0.78 $\pm$ 0.40 & 1.61 $\pm$ 0.41  & 2.08 $\pm$ 0.32 \\
H$_{\gamma}$ & 1.56 $\pm$ 0.17 & 4339.02 $\pm$ 0.60 & -1.44 $\pm$ 0.60 & 1.57 $\pm$ 0.33 & 1.81 $\pm$ 0.28 \\
H$_{\delta}$ & 1.45 $\pm$ 0.25 & 4099.08 $\pm$ 0.70 & -2.65 $\pm$ 0.70 & 1.56 $\pm$ 0.21 & 1.69 $\pm$ 0.19 \\
H$_{\epsilon}$ & 1.36 $\pm$ 0.27 & 3961.59 $\pm$ 0.80 & -8.48 $\pm$ 0.80 & 2.12 $\pm$ 0.10 & 2.18 $\pm$ 0.10\\
H$_{\zeta}$ & -- & 3875.31 $\pm$ 0.80 & -13.75 $\pm$ 0.80 & 2.11 $\pm$ 0.06 & 2.15 $\pm$ 0.06 \\
  & & & & \\
mean & 1.58 $\pm$ 0.08 & & & 2.06 $\pm$ 0.05 & 2.12 $\pm$ 0.05 \\
\end{tabular}
\end{table*}

\section{Results}
\subsection{Magnetic field strength from the Zeeman effect}

The average spectrum of PG 2329+267, presented in figures 1 and 2,
shows the line splitting due to the linear Zeeman effect. A peculiar
feature exists in the H$_{\alpha}$ spectrum shown in the bottom of
figure 2; while the $\sigma^{-}$ component is apparently split into
two, no such splitting is seen on the $\sigma^{+}$ component. The
feature appears to be real but we can think of no explanation for such
an asymmetry between the two $\sigma$ components.

We calculated the flux weighted average field strength over the
visible hemisphere of the white dwarf (B$_{s}$), by measuring the
degree of wavelength splitting due to the linear Zeeman effect. The
results are shown in table 1, column 2. For all the triplets, the line
positions of the $\pi$ and $\sigma$ components and their corresponding
uncertainties were measured by eye. The weighted mean of the
measurements from all the lines is, B$_{s}$ = 1.58 $\pm$ 0.08 MG. This
combines a measurement of the field strength from the H$_{\alpha}$
line and the higher lines of the Balmer series even though they were
taken at different times. The mean for all measurements taken from the
discovery spectrum alone, i.e not including H$_{\alpha}$, is B$_{s}$ =
1.52 $\pm$ 0.10 MG

We also calculated B$_{s}$, from the blue shift resulting from the
quadratic Zeeman effect. The measured line center of the $\pi$
component of each triplet is shown in table 1, column 3, the shift
from the rest wavelength is shown in column 4 and the resultant
magnetic field strength is shown in column 5. Note that the
H$_{\alpha}$ line has a small redshift (+0.28 $\pm$ 0.10 \AA) from its
rest wavelength, due to the combination of the gravitational redshift
of the white dwarf and its intrinsic space motion dominating over the
smaller blueshift from the quadratic Zeeman effect. If we assume a
magnetic field strength of B$_{s}$ = 1.58 $\pm$ 0.08 MG as measured
from the linear Zeeman effect, the calculated quadratic Zeeman shift
at H$_{\alpha}$ will be -0.43 $\pm$ 0.04 \AA. Hence the total shift
due to the gravitational redshift and intrinsic space motion will be
+0.71 $\pm$ 0.11 \AA\, at H$_{\alpha}$ or 32.4 $\pm$ 5.0 km/sec. This
velocity was used to correct the quadratic Zeeman shifts of the other
spectral lines, the magnetic fields were recalculated and are shown in
table 1, column 6. The mean value of B$_{s}$, calculated from the
uncorrected quadratic Zeeman shifts, is 2.06 $\pm$ 0.05 MG. Correcting
the shifts with an intrinsic redshift of 32.4 $\pm$ 5.0 km/sec
increases the measured value of B$_{s}$ to 2.12 $\pm$ 0.05 MG. Both
values are inconsistent with the value of B$_{s}$ determined from the
linear Zeeman effect. There is an increase in the determined value of
B$_{s}$ from the H$_{\epsilon}$ and H$_{\zeta}$ lines; the mean value
of B$_{s}$ for the three earlier lines H$_{\beta}$, H$_{\gamma}$ and
H$_{\delta}$ is 1.57 $\pm$ 0.16 MG when measured with the uncorrected
shifts and 1.79 $\pm$ 0.14 MG with the corrected shifts, which are
consistent with the mean value determined from the linear Zeeman
effect. To account for a possible systematic difference between the
H$_{\alpha}$ spectrum (taken with the WHT) and the H$_{\beta}$ -
H$_{\zeta}$ spectrum (taken with the INT) the magnetic field due to
the quadratic Zeeman effect was calculated using a range of different
velocity corrections (0 to 100 km/sec). No correction yields either a
self consistent set of field strengths from the INT spectrum or a mean
value consistent with the magnetic field strength measured from the
linear Zeeman effect. We believe this to be the result of the breakdown
of the perturbation theory used to derive equation 2.

\subsection{Magnetic field strength from circular spectropolarimetry}

The results of the circular spectropolarimetry are shown in figure
2. The bottom panel shows the normalised H$_{\alpha}$ flux while the
top panel shows the percentage of circular polarisation. The
polarisation spectrum clearly shows the S shaped profile indicative of
a magnetic field. The peak percentage of circular polarisation at
H$_{\alpha}$ is approximately 10\%. The longitudinal magnetic field
strength B$_{e}$ is calculated using equation 3, by a point-by-point
technique (Schmidt et al 1992), where in this instance the observed
profile is fitted by multiple Gaussians and the flux gradient is
calculated from this smooth profile in order to minimise the effect
that noise in the line profile has on the magnetic field
measurement. The calculated magnetic field strength is then a weighted
integral of the point-by-point measurements, made across the line
profile. We determined a value of B$_{e}$ = + 462 $\pm$ 60 KG, though
some caution should be taken with this figure as the weak field
approximation used in the equation will be breaking down as the Zeeman
components are resolvable. Observations of the spectropolarimetric
standard 53 Cam were used to obtain the correct sign for the magnetic
field, (Angel, Mcgraw $\&$ Stockman 1973). We follow the convention
that a positive circular polarisation corresponds to counterclockwise
rotation of the electric vector as seen by the observer.

\section{Discussion}

\subsection{Magnetic field strength and orientation of PG 2329+267}

We have calculated the mean magnetic field strength over the visible
hemisphere to be B$_{s}$ = 1.58 $\pm$ 0.08 MG. There is no sign of
rotational modulation of the magnetic field strength from our data as
the measured values from the Zeeman splitting of the H$_{\alpha}$
(B$_{s}$ = 1.47 $\pm$ 0.07 MG) and the other lines (B$_{s}$ = 1.52
$\pm$ 0.10 MG), which were taken on two separate occasions, are
consistent with each other.

The mean longitudinal magnetic field strength has been determined to
be B$_{e}$ = + 462 $\pm$ 60 KG, hence the ratio of the
longitudinal-to-mean field strength B$_{e}$/B$_{s}$ = 0.29 $\pm$
0.04. This can be used to place limits on the orientation at which we
observe the magnetic field. The longitudinal component of the magnetic
field will be largest at the magnetic pole where $i = 0^{\circ}$ and
will decrease as we look closer to the magnetic equator (at $i =
90^{\circ}$). We have assumed a simple centered dipole structure and
calculated that we are observing the magnetic field at an inclination
of $i = 60^{\circ}$ $\pm$ 5$^{\circ}$ from the magnetic axis. This
result is consistent with a comparison of the H$_{\alpha}$ spectrum in
figure 2 with computed spectra for a 3MG dipole field (Achilleos \&
Wickramasinghe 1989) which suggests that we must be observing the
magnetic field at an angle greater than 45$^{\circ}$. These limits
constrain the dipole magnetic field strength B$_{d}$ so that B$_{d}$ =
2.31 $\pm$ 0.59 MG, where the dipole field strength is calculated
using B$_{e}$ = 0.4B$_{d}$cos$i$ (Schmidt \& Smith 1995). This is a
relatively weak magnetic field and gives PG 2329+267 the fourth lowest
magnetic field strength of the 42 magnetic white dwarfs in Schmidt and
Smith's (1995) list.

\subsection{The mass of PG 2329+267}

The discovery spectrum taken with the INT is shown in figure 1 along
with two comparison, non-magnetic white dwarf spectra, which were
taken with an identical setup. WD 1134+300 is a white dwarf with a
mass of 0.9M$_{\odot}$ (Bergeron et al 1992) and an effective
temperature of 14,000 K, similar to that of PG 2329+267 for which the
effective temperature is approximately 10,000 K (Shipman 1979). With
the exception of the Zeeman features these two spectra are remarkably
similar, particularly in the number of visible Balmer lines, which
suggests that they have similar masses. For further comparison the
spectrum of WD 1344+572 is shown in the bottom of figure 1. This white
dwarf has a mass of 0.56M$_{\odot}$, at the peak of the white dwarf
mass distribution (Bergeron et al 1992), and a temperature of 21,700
K. The increased mass of WD 1134+300 and PG 2329+267 compared to WD
1344+572 is evident from the smaller number of visible Balmer
lines. The higher surface gravity of a more massive white dwarf
increases the Stark broadening of the absorption lines of the hydrogen
Balmer series and also reduces the number of visible absorption
lines. By comparing the number of the higher Balmer lines visible in
the spectra of the three white dwarfs we can deduce that WD 2329+267
is more massive than the majority of white dwarfs and may have a mass
comparable to WD 1134+300 at 0.9M$_{\odot}$.

\subsection{The evolution of PG 2329+267}

If we consider the origin of the magnetic field in PG 2329+267 we can
find evidence both for and against the hypothesis that it evolved from
a chemically peculiar, magnetic Ap or Bp star. The magnetic field
strength of PG 2329+267 is consistent with the theory of magnetic flux
conservation from a magnetic main sequence Ap or Bp star. As we showed
with a simple qualitative argument we believe PG 2329+267 to be more
massive than the majority of white dwarfs, suggesting it evolved from
a fairly massive progenitor such as an Ap or Bp star. However, if we
consider the galactic space motion of PG 2329+267, which is given as
85.8 km/sec (Sion et al 1988), we can see it is much larger than that
found for all other magnetic white dwarfs, which themselves form a
distinct low velocity kinematic sub-group (Sion et al 1988). The
magnetic white dwarfs considered by Sion et al, with the exception of
WD 0912+536, have low velocities with respect to the sun ($<$50km/sec)
indicating their youth and evolution from massive progenitors, such as
Ap and Bp stars. This is however only a statistical argument and we
don't consider the high velocity of PG 2329+267 to necessarily rule
out its evolution from an Ap or Bp main sequence star, which remains
the most plausible hypothesis for the production of magnetic white
dwarfs.

\section{Conclusions}

We have detected a magnetic field from the white dwarf PG
2329+267. The mean surface field strength measured from the degree of
linear Zeeman splitting of the Balmer hydrogen lines is B$_{s}$ = 1.58
$\pm$ 0.08 MG. Similar measurements from the quadratic Zeeman effect
yield consistent results only for the Balmer lines up to H$_{\delta}$,
for higher lines the perturbation theory used to calculate the
magnetic field strength begins to break down. We have detected
approximately 10\% circular polarisation at H$_{\alpha}$ and have
calculated the mean longitudinal magnetic field strength to be B$_{e}$
= + 462 $\pm$ 60 KG. The ratio B$_{e}$/B$_{s}$ = 0.29 $\pm$ 0.04 and
the shape of the Zeeman split components suggest we are viewing the
white dwarf at an inclination of $i = 60^{\circ}$ $\pm$ 5$^{\circ}$
from the magnetic axis. At this inclination the dipole magnetic field
strength will be 2.31 $\pm$ 0.59 MG making PG 2329+267 the fourth
weakest known isolated magnetic white dwarf. We have suggested that PG
2329+267 is more massive than most isolated white dwarfs which
supports the hypothesis that magnetic white dwarfs evolve from
chemically peculiar main sequence Ap and Bp stars.

\subsection{Acknowledgments}
 
This work was based on observations from the INT and WHT operated on
the island of La Palma by the Isaac Newton Group in the Spanish
Observatorio del Roque de los Muchachos of the Instituto de
Astrofisica de Canarias. C. Moran was supported by a PPARC studentship
and TRM was supported by a PPARC Advanced Fellowship.

\bsp

\end{document}